# Probing the Circumgalactic Medium with Fast Radio Bursts: Insights from CAMELS

Isabel Medlock[1], Daisuke Nagai[1,2], Priyanka Singh[2,3], Benjamin Oppenheimer[4], Daniel Anglés-Alcázar[5,6], and Francisco Villaescusa-Navarro[6,7]
[1] Department of Astronomy, Yale University, New Haven, CT 06520, USA; isabel.medlock@yale.edu
[2] Department of Physics, Yale University, New Haven, CT 06520, USA
[3] Department of Astronomy, Astrophysics and Space Engineering, Indian Institute of Technology Indore, Khandwa Road, Simrol, 453552, India
[4] CASA, Department of Astrophysical and Planetary Sciences, University of Colorado, 389 UCB, Boulder, CO 80309, USA
[5] Department of Physics, University of Connecticut, 196 Auditorium Road, U-3046, Storrs, CT 06269, USA
[6] Center for Computational Astrophysics, Flatiron Institute, 162 5th Avenue, New York, NY 10010, USA
[7] Department of Astrophysical Sciences, Princeton University, Peyton Hall, Princeton, NJ 08544, USA

## Abstract

Most diffuse baryons, including the circumgalactic medium (CGM) surrounding galaxies and the intergalactic medium (IGM) in the cosmic web, remain unmeasured and unconstrained. Fast radio bursts (FRBs) offer an unparalleled method to measure the electron dispersion measures (DMs) of ionized baryons. Their distribution can resolve the missing baryon problem and constrain the history of feedback theorized to impart significant energy to the CGM and IGM. We analyze the Cosmology and Astrophysics with Machine Learning Simulations using three suites, IllustrisTNG, SIMBA, and Astrid, each varying six parameters (two cosmological and four astrophysical feedback), for a total of 183 distinct simulation models. We find significantly different predictions between the fiducial models of the suites owing to their different implementations of feedback. SIMBA exhibits the strongest feedback, leading to the smoothest distribution of baryons and reducing the sight-line-to-sight-line variance in DMs between $z = 0$ and 1. Astrid has the weakest feedback and the largest variance. We calculate FRB CGM measurements as a function of galaxy impact parameter, with SIMBA showing the weakest DMs due to aggressive active galactic nucleus (AGN) feedback and Astrid the strongest. Within each suite, the largest differences are due to varying AGN feedback. IllustrisTNG shows the most sensitivity to supernova feedback, but this is due to the change in the AGN feedback strengths, demonstrating that black holes, not stars, are most capable of redistributing baryons in the IGM and CGM. We compare our statistics directly to recent observations, paving the way for the use of FRBs to constrain the physics of galaxy formation and evolution.

*Unified Astronomy Thesaurus concepts:* Circumgalactic medium (1879); Radio transient sources (2008); Hydrodynamical simulations (767); Stellar feedback (1602)

## 1. Introduction

The circumgalactic medium (CGM) refers to the area beyond the galactic disk and the interstellar medium (ISM) but within the virial radius of the halos. The CGM serves as a reservoir of diffuse gas and plasma and can contain up to 80% of the baryonic mass within the dark matter halo (Anderson & Bregman 2010; Peeples et al. 2013; Werk et al. 2014; Tumlinson et al. 2017). This makes the baryonic content of the CGM a partial solution to the missing baryon problem, where we only observe about 5%–10% of the expected baryons within the stellar and ISM content of galaxies (Cen & Ostriker 1999; Bregman 2007; Shull et al. 2012). The dynamics of the CGM are complex and governed by various feedback processes, including gas cooling, supernova feedback, and active galactic nuclei (AGN) feedback (e.g., Anderson & Bregman 2010; Naab & Ostriker 2017). The CGM receives gas from inflows from the intergalactic medium (IGM) and winds from central and satellite galaxies, which in turn fuels star formation (e.g., Oppenheimer et al. 2010; Christensen et al. 2016; Anglés-Alcázar et al. 2017b). Understanding these feedback processes and their interface with the CGM is crucial in understanding the formation and evolution of galaxies and clusters.

Recent advances in galaxy formation modeling and simulations have enabled detailed modeling of baryonic feedback processes (Somerville & Davé 2015; Vogelsberger et al. 2020). To study the CGM, cosmological simulations have been used to examine the structure of the CGM (e.g., Peeples et al. 2019; van de Voort et al. 2019) and determine the origin of the gas mass (e.g., Hafen et al. 2019). However, as it is challenging to model the wide range of galactic scales computationally, subgrid models are needed in simulations (e.g., Genel et al. 2014; Anglés-Alcázar et al. 2017a). The implementation of these models can lead to significantly different predictions, particularly for the matter power spectrum (e.g., Delgado et al. 2023; Gebhardt et al. 2024) and the distribution of gas around galaxies, which is highly dependent on feedback models (e.g., Chisari et al. 2018; van Daalen et al. 2020). The cold gas of the ISM is also highly sensitive to subgrid models, and even variations within models, as shown in the comparison of SIMBA, EAGLE, and IllustrisTNG in Davé et al. (2020).

Despite these challenges, there are exciting developments taking place in observational astronomy. Rapidly improving observations with greater sensitivity and resolution allow us to detect and study elusive CGM gas. Crucially, multiwavelength astronomical studies enable us to use a variety of probes spanning ranges of mass and redshift (Battaglia et al. 2019). Recently, strides have been made in observing CGM through thermal and kinetic Sunyaev–Zel'dovich (SZ) effects (e.g., Amodeo et al. 2021; Bregman et al. 2022) as well as through X-ray emission with eROSITA (e.g., Chadayammuri et al. 2022; Comparat et al. 2022).

One promising area of study is the discovery of fast radio bursts (FRBs), first observed by Lorimer et al. (2007). FRBs are millisecond long luminous extragalactic pulses of radio waves of uncertain origin (Petroff et al. 2019). Although the origins of FRBs are not yet well understood, several theories have been postulated as to possible progenitors involving magnetars (e.g., Platts et al. 2019). The discovery of four more bursts by Thornton et al. (2013) confirmed the discovery of a new phenomenon. Since then, facilities such as CHIME, the Parkes telescope, and ASKAP have led to an exponential increase in FRB detections (e.g., Bhandari et al. 2018; Shannon et al. 2018; CHIME/FRB Collaboration et al 2021).

FRBs are not only an intriguing phenomenon, they are also valuable for cosmological research. One of their primary observables, the dispersion measure (DM), is crucial in these efforts. Since their DM is a proxy for the electron column density, FRBs are an excellent tool for tracing baryons (Fujita et al. 2017; Battaglia et al. 2019; Ravi 2019). For example, DMs of FRB populations can help locate and measure missing baryons, as demonstrated by studies such as McQuinn (2014), Muñoz & Loeb (2018), and Macquart et al. (2020). FRB foreground mapping, as performed with the FLIMFLAM survey, can be used to constrain the cosmic baryon distribution in the cosmic web and the partition between the IGM and the CGM (Simha et al. 2020, 2021, 2023; Lee et al. 2022, 2023). In addition, FRBs show potential as a probe of the CGM (e.g., McQuinn 2014; Ravi et al. 2019). As most of the gas is ionized, the CGMs of the foreground halos add a significant excess DM to the total DM of an FRB. Recent efforts have been made to measure this excess DM with the CHIME catalog (Connor & Ravi 2022; Wu & McQuinn 2023).

These studies rely on a large population of well-localized FRBs, that is, FRBs with precise enough spatial and redshift localization to map to a host halo. Consequently, especially exciting is the increase in precise localizations of FRBs, thanks to surveys and facilities such as CHIME, ASKAP, and DSA-110 (CHIME/FRB Collaboration et al 2018; Macquart et al. 2020; Law et al. 2023). The number of localizations is expected to increase with new facilities such as CHIME outriggers, CHORD, and DSA-2000, which aims to simultaneously detect and localize ∼10,000 FRBs per year (Hallinan et al. 2019; Vanderlinde et al. 2019; Leung et al. 2021), sufficient to access and characterize baryon contents and physical conditions in hot and diffuse CGM, ICM, and IGM (Ravi et al. 2019).

In this paper, we examine the potential of using FRBs to constrain the subgrid physical models that shape the CGM in hydrodynamical simulations. The paper is structured as follows. First, in Section 2, we review the necessary background information on FRBs, their DMs, and the $F$ parameter. In Section 3, we will describe the Cosmology and Astrophysics with Machine Learning Simulations (CAMELS) used and the steps we took to perform our analysis. In Section 4, we discuss our findings on cosmic DM, the contribution of CGM to DM, and the effect of feedback as quantified by the $F$ parameter. In Section 5, we discuss the limitations of our work and possible next steps to address these. Finally, in Section 6, we briefly summarize our work and list our conclusions.

## 2. FRBs and DM

When we detect an FRB, we observe a sharp pulse (similar to that of a pulsar) with dispersion caused by the medium through which it passes. As the radio pulse travels from the FRB source to Earth through an ionized medium, the photons interact with charged particles, causing a time delay in the propagation of the signal. The time delay is inversely proportional to the mass of the charged particle, with electrons having a more significant effect than protons. This time delay is also a function of the wavelength/frequency of the signal, where more energetic photons experience less of an effect. Thus, when we detect an FRB, we measure the time delay between the highest and lowest radio frequencies of observation, following the equation

$$\Delta t = \frac{e^2}{2\pi m_e c}(\nu_{\mathrm{lo}}^{-2} - \nu_{\mathrm{hi}}^{-2})\,\mathrm{DM}, \quad (1)$$

where $\nu_{\mathrm{lo}}$ and $\nu_{\mathrm{hi}}$ are the lowest and highest frequency of observation, respectively; $m_e$ is the mass of an electron; and $c$ is the speed of light (Petroff et al. 2019). Thus, as the signal travels from the source and traverses the intervening medium to Earth, it directly traces ionized baryons that cannot be detected using other observational methods.

The DM is equal to the integrated electron density along the line of sight (LOS) from the source to the observer, mathematically defined as

$$\mathrm{DM} = \int_0^d \frac{n_{\mathrm{e}}(l)}{1+z}dl, \quad (2)$$

where $d$ is the proper distance to the source, $n_{\mathrm{e}}$ is the physical free electron number density, $z$ is the redshift, and $l$ is the proper path length. DM is given in units of electron number density over path length rather than in units of column density.

The signal from an FRB experiences dispersion as it travels through the Universe before reaching an observer due to the Universe's ionized components. The total DM is the sum of these components:

$$\mathrm{DM_{obs}} = \mathrm{DM_{MW}} + \mathrm{DM_{IGM}} + \mathrm{DM_{CGM}} + \frac{DM_{\mathrm{Host}}}{1+z}, \quad (3)$$

where $\mathrm{DM_{MW}}$ represents the amount of dispersion contributed by the Milky Way, $\mathrm{DM_{IGM}}$ is the contribution from the IGM, $\mathrm{DM_{CGM}}$ denotes the contribution from the CGM of halos that the sight line may intersect, and $\mathrm{DM_{Host}}$ is the contribution from the host of the source itself, which is scaled by its redshift to account for dispersion in the host galaxy's rest frame, following the convention established in Macquart et al. (2020).

Much effort has been made in the estimation of $\mathrm{DM_{IGM}}$, both analytically and via cosmological simulations (Ioka 2003; Inoue 2004; McQuinn 2014; Macquart et al. 2020; Zhu & Feng 2021). For example, Deng & Zhang (2014) derive the contribution of the IGM to be

$$\langle \mathrm{DM_{IGM}} \rangle = A \int_0^z \frac{(1+z)x(z)}{\sqrt{\Omega_M(1+z)^3 + \Omega_\Lambda}}dz, \quad (4)$$

where $A$ is a constant equal to 933 cm$^{-3}$ pc (for the standard Planck cosmological parameters used), $x(z)$ is the ionization fraction function, $\Omega_M$ is the energy density of matter, and $\Omega_\Lambda$ is the cosmological constant.

The Macquart relation is a well-known formula that defines $DM_{cosmic} = DM_{IGM} + DM_{CGM}$ as a function of the redshift. The value of $DM_{cosmic}$ takes into account the contribution of the IGM and any intersecting halos but does not include the host or the Milky Way (Macquart et al. 2020). This relation is defined as

$$\langle DM_{cosmic} \rangle = \int_0^{z_{FRB}} \frac{c f_d \rho_b(z) m_p^{-1} (1 - Y_{He}/2)}{H_0 (1+z)^2 \sqrt{\Omega_M (1+z)^3 + \Omega_\Lambda}} dz, \quad (5)$$

where $f_d$ is the fraction of cosmic baryons in diffuse ionized gas, $\rho_b(z) = \Omega_b \rho_{c,0} (1+z)^3$, $m_p$ is the proton mass, $Y_{He} = 0.25$ is the mass fraction of helium assumed doubly ionized, and $\Omega_M$ and $\Omega_\Lambda$ are the total matter and dark energy densities at the present day in units of the critical density.

Another major contributor to the DM is the Milky Way, for which a large uncertainty remains. The Milky Way's contribution to DM ($DM_{MW}$) can be divided into two parts: $DM_{MW,ISM}$ and $DM_{MW,Halo}$. $DM_{MW,ISM}$ can be modeled following the NE2001 (Cordes & Lazio 2002) and YMW16 (Yao et al. 2017) Galactic plasma models, partially based on pulsar DMs with independent distances. The contribution of the Milky Way halo is estimated to be $DM_{MW,Halo} = 10-80 \text{ cm}^{-3}$ pc (Prochaska & Zheng 2019; Keating & Pen 2020). In previous work, $DM_{MW,Halo}$ is estimated to be 50 cm$^{-3}$ pc (e.g., Macquart et al. 2020; James et al. 2022). Recent observations have measured upper limits of $DM_{MW,Halo} = 28.7/47.3 \text{ cm}^{-3}$ pc[8] (Ravi et al. 2023) and $DM_{MW,Halo} = 52-111 \text{ cm}^{-3}$ pc (Cook et al. 2023).

It is difficult to accurately characterize $DM_{Host}$ due to its large uncertainty and the lack of well-localized FRBs. To model $DM_{host}$, the following lognormal probability distribution is often used:

$$p(DM'_{host}) = \frac{1}{DM'_{host}} \frac{1}{\sigma_{host} \sqrt{2\pi}} \exp\left[-\frac{(\log DM'_{host} - \mu_{host})^2}{2\sigma_{host}^2}\right], \quad (6)$$

where $DM'_{host}$ is the host halo DM contribution scaled by redshift ($DM_{host}/(1+z)$), and $\mu_{host}$ and $\sigma_{host}$ are free parameters (Macquart et al. 2020).

The presence of ionized CGM gas causes FRBs passing through to experience an increase in DM. However, accurately characterizing the DM of the CGM requires subtracting the contributions from both the Milky Way and the IGM. To achieve this, precise measurements of the FRB's spatial position with arcsecond or subarcsecond localizations are necessary. If the spatial location of the FRB overlaps with a galaxy of known redshift, we can attribute a redshift to the FRB, assuming that the galaxy is indeed the host (see, e.g., Aggarwal et al. 2021). This allows us to subtract the contribution of the host galaxy to the DM and identify any intervening galaxies responsible for excess DM.

## 3. Methods

### 3.1. CAMELS

The CAMELS[9] project consists of 10,680 simulations, including 5516 magnetohydrodynamic simulations and 5164 N-body simulations (Villaescusa-Navarro et al. 2021a, 2023).

CAMELS comprises different simulation suites using different models: IllustrisTNG, SIMBA, Astrid, Magneticum, SWIFT-EAGLE, Ramses, Enzo, and N-body. The N-body suite has a corresponding dark-matter-only simulation for each CAMELS hydrodynamic simulation, with the same cosmology and random seed value, and is run with Gadget-3 (Springel 2005). Currently, only SIMBA, IllustrisTNG, and Astrid are available for public analysis. CAMELS-IllustrisTNG has 2143 hydrodynamic simulations run with the AREPO code (Springel 2010; Weinberger et al. 2020) and the same subgrid physics as the original IllustrisTNG simulations (Weinberger et al. 2017; Pillepich et al. 2018). CAMELS-SIMBA has 1092 hydrodynamic simulations run with the GIZMO code (Hopkins 2015) and the same subgrid physics as the original SIMBA simulations (Davé et al. 2019). CAMELS-Astrid has 2116 hydrodynamic simulations run with the MP-Gadget code, which is a highly scalable version of Gadget-3 (Springel 2005). The galaxy formation subgrid model is described in Ni et al. (2022) and is based on the original Astrid code described in Bird et al. (2022). It is essential to note that the three suites differ in their implementation of gravity and hydrodynamic solvers, radiative cooling parameterization, star formation and evolution, and feedback from galactic winds and AGN. For more information, we refer to Ni et al. (2023).

All simulations follow the evolution of $256^3$ dark matter particles and fluid elements (for hydrodynamical simulations) from $z = 127$ to $z = 0$ in a periodic box with sides of comoving length $L = 25 \, h^{-1}$ Mpc. The gas mass resolution is initially set to $1.27 \times 10^7 \, h^{-1} \, M_\odot$. Dark matter resolution elements are set to masses of $6.49 \times 10^7 (\Omega_M - \Omega_b)/0.251 h^{-1} \, M_\odot$. All simulations are run with the following cosmological parameters: $\Omega_b = 0.049$, $h = 0.6711$, $n_s = 0.9624$, $\Sigma m_s = 0.0$ eV, and $w = -1$.

To analyze the impact of each parameter individually, we utilized the one parameter at a time (1P) set of each CAMELS suite, which consists of 61 simulations for each suite. In this set, only one cosmological or astrophysical parameter is varied at a time (hence the name 1P set), while the random seed values (initial conditions) remain constant.

The 1P set has two cosmological parameters ($\Omega_M$ and $\sigma_8$) and four astrophysical parameters, which are related to stellar feedback ($A_{SN1}$ and $A_{SN2}$) and AGN feedback ($A_{AGN1}$ and $A_{AGN2}$). $\Omega_M$ is the fraction of energy density in matter (baryonic and dark matter combined) and is linearly sampled from 0.1 to 0.5, with the fiducial value set at 0.30. $\sigma_8$ is the variance of the linear field on a scale of $8 \, h^{-1}$ Mpc at $z = 0$ and is linearly sampled from 0.6 to 1.0, with the fiducial value set at 0.80. Table 1 summarizes key information on the four astrophysical parameters used in this study for the IllustrisTNG, SIMBA, and Astrid models. The feedback parameter values range as listed in the bottom row, where the fiducial value of 1.0 corresponds to the value used in the original simulation. The physical meanings of these parameters vary depending on the suite/subgrid model, which is summarized in Table 1 for each suite.

One of the primary goals of CAMELS is to enhance the scientific output of current and future missions such as DESI, eROSITA, the Roman Observatory, and the Rubin Observatory. To achieve this goal, we require reliable theoretical predictions that factor in the uncertainties associated with our limited understanding of baryonic feedback processes. The CAMELS project has made significant progress in this regard. For example, Villaescusa-Navarro et al. (2021b) used

---

[8] This estimate depends on which of the two nearby pulsars is used to estimate $DM_{ISM}$.
[9] https://www.camel-simulations.org/

**Table 1**
Physical Meanings of the Four Astrophysical Parameters for the IllustrisTNG, SIMBA, and Astrid Suites of CAMELS

| Simulation | $A_{SN1}$ | $A_{SN2}$ | $A_{AGN1}$ | $A_{AGN2}$ |
|---|---|---|---|---|
| IllustrisTNG | Galactic winds: energy per unit SFR | Galactic winds: wind speeds | Kinetic mode BH feedback: energy per unit BH accretion rate | Kinetic mode BH feedback: ejection speed/burstiness |
| SIMBA | Galactic winds: mass loading | Galactic winds: wind speeds | QSO and jet mode BH feedback: momentum flux | Jet mode BH feedback: jet speed |
| Astrid | Galactic winds: energy per unit SFR | Galactic winds: wind speeds | Kinetic mode BH feedback: energy per unit BH accretion rate | Thermal mode BH feedback: energy per unit BH accretion rate |
| Variation range | [0.25–4.0] | [0.5–2.0] | [0.25–4.0] | [0.5–2.0] |

**Note.** Note that these physical meanings vary depending on the subgrid model used. The range of values for each of the four parameters is shown in the last row. The only exception is for the Astrid suite $A_{AGN2}$ parameter, which varies from 0.25 to 4.0, not from 0.5 to 2.0, as stated in the table. The parameters are logarithmically sampled in the specified ranges, and a value of 1.0 corresponds to that used in the original simulation.

**Table 2**
CAMELS Snapshots and Corresponding Redshift Used in This Analysis

| z | 0.05 | 0.27 | 0.54 | 0.95 | 1.48 | 2.00 |
|---|------|------|------|------|------|------|
| SIMBA | 032 | 028 | 024 | 019 | 014 | 010 |
| IllustrisTNG | 032 | 028 | 024 | 019 | 014 | 010 |
| Astrid | 088 | 080 | 072 | 062 | 052 | 044 |

CAMELS to robustly constrain $\Omega_M$ and $\sigma_8$ to a percent level with 2D maps of the total matter mass of the hydrodynamical simulations while marginalizing over uncertainties in baryonic physics. Related to our work, Nicola et al. (2022) used CAMELS to investigate the potential use of the auto-power spectrum and the cross-power spectrum of the baryon distribution to constrain baryonic feedback and cosmology. With its diverse range of cosmological and feedback models, as well as subgrid model-based code suites, CAMELS is the ideal tool to investigate the impact of each of these on the CGM using FRBs as a probe. For more information on the CAMELS project, we refer the reader to Villaescusa-Navarro et al. (2021a, 2023) and Ni et al. (2023).

### 3.2. DM Maps and Halo Catalogs

We create DM maps from the CAMELS 1P set for the snapshots and the corresponding redshifts listed in Table 2. For Astrid, SIMBA, and IllustrisTNG, we take these snapshots and calculate electron column density maps from the electron density field with yt (Turk et al. 2011). yt interpolates the electron density contained in all fluid elements onto 2D grids, effectively collapsing the entire CAMELS volume depth. This grid is $4000 \times 4000$ pixels with a pixel size of 9.3 comoving kpc across and a depth of $l = 25$ comoving $h^{-1}$ Mpc. Each pixel represents the electron column density throughout the volume. We account for the redshift by dividing the results by $(1 + z)$ as described in Equation (2).

For reference, Figure 1 shows the halo mass distributions of the top 300 most massive halos for the fiducial models of SIMBA, IllustrisTNG, and Astrid for the $z = 0.05$ snapshot. These halos fall within a mass range of $\log(M/M_\odot) \in [11.0, 13.5]$. SIMBA and Astrid exhibit similar distributions in the halo mass, with a more pronounced peak near $M = 10^{11.6} M_\odot$ compared to IllustrisTNG.

Significant differences exist in the SIMBA, IllustrisTNG, and Astrid DM maps. The SIMBA model shows that AGN feedback can distribute baryons to larger scales, up to 12 $h^{-1}$ Mpc in extreme cases (Davé et al. 2019; Borrow et al. 2020; Gebhardt et al. 2024). In some instances, the AGN jet feedback might even cause up to 80% of baryons to evacuate halos by the present time (Appleby et al. 2021; Sorini et al. 2022). Compared to the IllustrisTNG and Astrid maps, the filaments in the SIMBA map are more diffuse. Additionally, smaller IllustrisTNG and Astrid halos have a higher DM compared to their SIMBA counterparts.

### 3.3. Building Light Cones

FRBs are of extragalactic origin and cosmologically distant. As each simulation cube has a comoving length of $l = 25$ $h^{-1}$ Mpc, it is necessary to stack the boxes to achieve sight lines out to cosmic distances. We calculate the running sum of the DM of each LOS by stacking boxes to the desired redshift and the corresponding distance. We choose our sight lines to be parallel to the simulation box edges, following convention (Jaroszynski 2019; Batten et al. 2021; Zhang et al. 2021). In addition, we keep a running sum of the cosmological distance our LOS has traversed. For each step, we convert the current distance of our LOS to a redshift $z_{\text{mean}}$, where we set $z_{\text{mean}}$ equal to the redshift of our current distance plus half a box length, the average redshift of the box step. We then determine which of the available redshift snapshot DM maps is closest to $z_{\text{mean}}$. Having selected our map, we randomly selected a column on the map to integrate over to ensure that we are not repeating over the same structures. We take this DM value and account for the redshift by dividing by $(1 + z_{\text{mean}})$. We update our running sum of the distance using a weighted box length, which we calculate by multiplying the box length ($l = 25$ $h^{-1}$ Mpc) by a factor of $((1 + z_{\text{mean}})/(1 + z_{\text{snapshot}}))^2$. We compute the sight lines for $z \in [0.1, 0.25, 0.5, 0.75, 1.0]$.

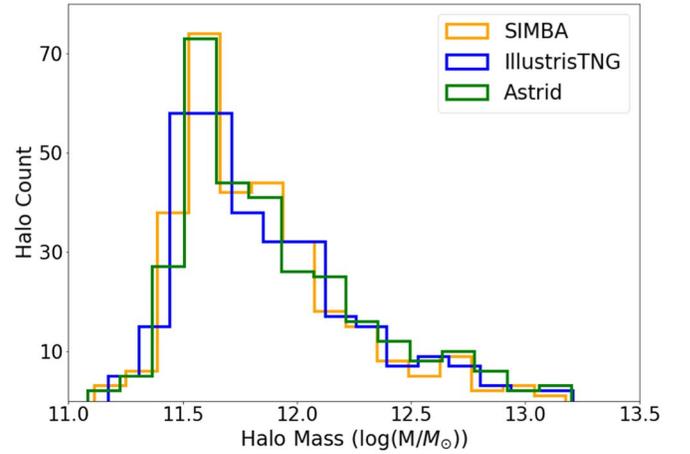

**Figure 1.** Halo mass distributions of the top 300 most massive halos for the fiducial $z = 0.05$ snapshots of SIMBA (orange), IllustrisTNG (blue), and Astrid (green).

Figure 2 displays the DM maps for the SIMBA, IllustrisTNG, and Astrid fiducial runs at $z = 0.05$. The map also includes the locations of the halos, represented by red dots. The halo catalogs are SUBFIND (Dolag et al. 2009) outputs run on all CAMELS, and we show the position of the central subhalo defined by the location of its central galaxy. The DM distributions corresponding to the fiducial model for SIMBA, IllustrisTNG, and Astrid are shown in Figure 3. The left panel displays the distributions over all cells of a single box, while the right panel shows the distributions of sight lines integrated out to $z = 1$. To calculate single-box distributions, we simply take into account all sight lines in the $4000 \times 4000$ pixel box. In the case of the integrated DM, we repeat the light-cone building process for 10,000 sight lines and analyze the resulting distribution. In Figure 4, we show the DM distributions of the integrated sight lines to $z = 1$ that characterize the entirety of the 1P set.

We note that there are a couple of limitations and uncertainties in this procedure. First, the length of the CAMELS box ($l = 25$ $h^{-1}$ Mpc) is not sufficiently large to fully capture cosmic variance due to the large-scale structure. We discuss this in more detail in Section 5. In addition, the LOS building procedure introduces uncertainty. The CAMELS project outputs a limited set of redshift snapshots, which requires us to create interpolated maps as described previously, similar to the procedure of Batten et al. (2021). It is common to perform a sequence of mirrors, rotations, and translations of simulation boxes to minimize sight lines intersecting through the same small structure. This is ineffective in the boxes of 25 and 50 $h^{-1}$ Mpc (Batten et al. 2021). Instead, we use a scrambling technique (randomly and independently selecting columns for each new box) when selecting the LOS through each stacked box, which minimizes the correlations of small structure between maps. These technicalities may lead to slight differences in the prediction of DM distributions between simulations (Walker et al. 2024).

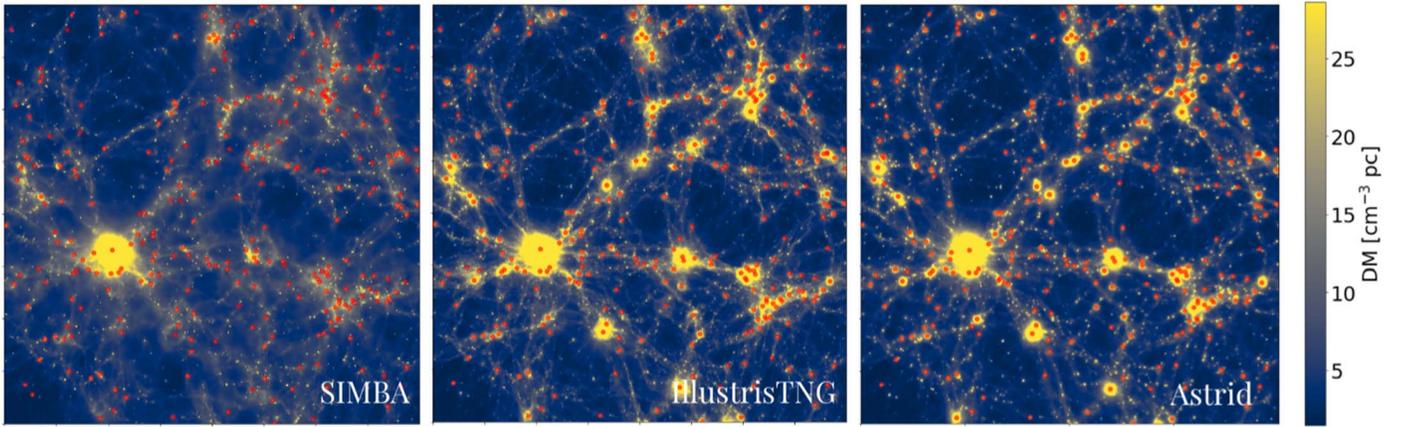

**Figure 2.** The DM maps over one CAMELS box ($L = 25\ h^{-1}$ Mpc) of the fiducial run of each of the three suites: SIMBA (left), IllustrisTNG (center), and Astrid (right). The snapshot corresponds to $z = 0.05$, and the locations of the 300 most massive halos at that redshift in each simulation are denoted by red points. The procedure for generating the DM maps is described in detail in Section 3.2. The color bar on the right maps the DM values to each cell in the map, where the halos are oversaturated and can reach values of up to 1000 cm$^{-3}$ pc. Moving from left to right, each suite shows a more concentrated distribution of baryons in and around halos, indicative of weaker feedback.

### 3.4. Halocentric Profiles

To assess the contribution of CGM to the total DM, we calculate the halo-mass-averaged DM profiles of halos as a function of their impact parameter. We have developed a method to isolate DM sight lines that intersect through a specific halo. We use the DM maps at $z = 0.05$ and the corresponding halo catalog in the profiles. To make the halocentric profiles, we follow the following steps. First, we map the halo location coordinates to the DM map coordinates. Then, for each cell on the DM map, we find the two closest halos and calculate the impact parameter ($b = R/R_{200}$). Using the impact parameter as our distance measure, we effectively normalize over the halo mass. Next, we discard any cell with $b < 3$ for the second-closest halo to minimize contamination by other halos in the profiles. Then, for each halo in the catalog, we select all DM cells for which this is the closest, bin the DM values by impact parameter, and take the median of all the DMs in each bin.

### 3.5. Calculating the F Parameter

Following the derivation of Macquart et al. (2020) based on calculations by McQuinn (2014), we can define a quantity referred to as the $F$ parameter to characterize the strength of feedback. Equation (5) describes the Macquart relation (i.e., DM$_{\text{cosmic}}$ as a function of redshift). The probability of deviation from the mean DM for a given $z$ is given by the probability distribution,

$$p_{\text{cosmic}}(\Delta) = A\Delta^{-\beta}\exp\left[-\frac{(\Delta^{-\alpha} - C_o)^2}{2\alpha^2\sigma_{\text{DM}}^2}\right],\quad \Delta > 0, \quad (7)$$

where $\Delta = \text{DM}_{\text{cosmic}}/\langle\text{DM}_{\text{cosmic}}\rangle$, $\alpha = \beta = 3$, $C_0$ is tuned so the expectation value is unity, and $\sigma_{\text{DM}}$ describes the spread of the distribution. This distribution provides excellent matches to both semianalytical models and hydrodynamic simulations (e.g., Macquart et al. 2020; Zhang et al. 2021). Cosmological simulations (e.g., the "swinds" simulations; Faucher-Giguère et al. 2011) show that the standard deviation of this distribution $\sigma_{\text{DM}}$ can be described as

$$\sigma_{\text{DM}}(\Delta) = Fz^{-1/2}, \quad (8)$$

where $\sigma_{\text{DM}}(\Delta)$ is the fractional standard deviation of the DM, $z$ is the redshift, and $F$ is the $F$ parameter. A decrease in the value of $F$ indicates an increase in feedback strength, while the value of $F$ approaching 1 indicates weaker feedback. Stronger feedback describes the situation where baryons are pushed out of halos, and weaker feedback refers to the situation where halos retain more of their baryons. The $z^{-1/2}$ scaling is expected due to the nearly Poisson nature of intersecting halos. Note that $F$ is sensitive to the overall distribution of baryons, including the large-scale structure that is not sensitive to feedback. To account for this, recently, $F$ has been dubbed the "fluctuating" parameter rather than the "feedback" parameter.

To calculate $F$ in the CAMELS 1P set, we consider 10,000 FRB sight lines out to a chosen redshift. For our analysis, we calculated the distributions at $z = 0.5$, since $F$ is expected to be constant in the redshift range of 0.4 to 2 (Zhang et al. 2021; Baptista et al. 2024). To calculate $\sigma_{\text{DM}}(\Delta)$ for each sight line, we take the standard deviation of the DM of the sight lines and divide it by the mean DM. Then, using Equation (8) and our chosen redshift ($z = 0.5$), we find $F$. We calculate the error in $F$ through bootstrapping analysis. Taking the 10,000 sight lines, we select a subset of 1000 and calculate $F$ based on this distribution. We repeat this process 10,000 times, randomly selecting a new subset. This results in an array with 10,000 $F$ values for the simulation from which we can calculate the standard deviation and confidence intervals. We calculate $F$ for each simulation in the 1P set of SIMBA, Astrid, and IllustrisTNG. To confirm that our calculations are consistent within the redshift range where $F$ is expected to be constant, we calculate $F$ at $z = 0.75$.

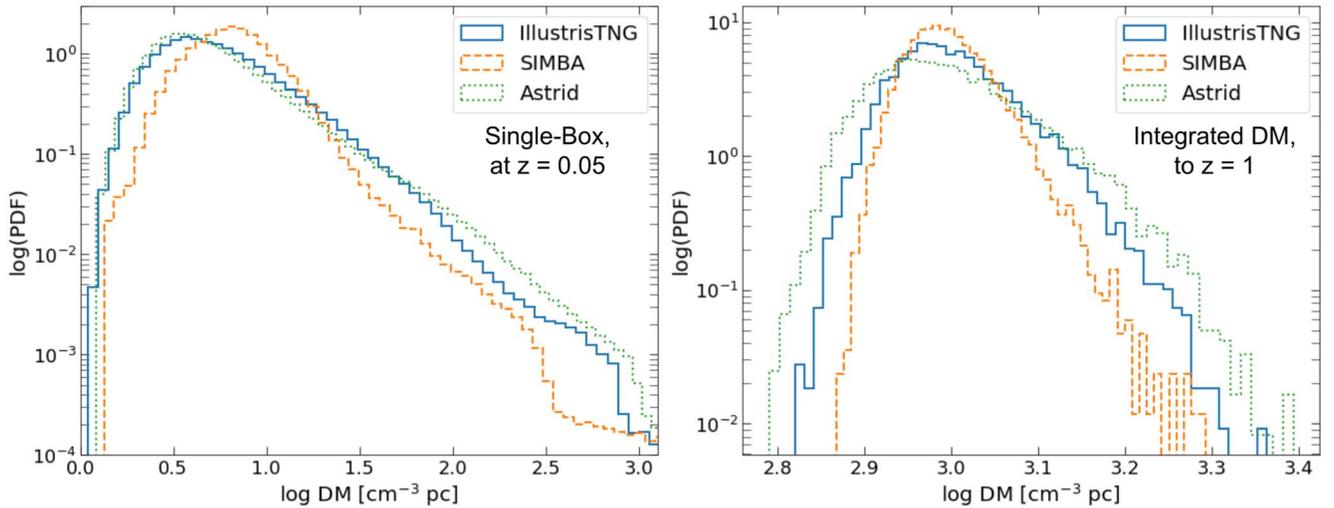

**Figure 3.** DM distributions corresponding to the fiducial model for SIMBA, IllustrisTNG, and Astrid. The left panel shows the distributions over all cells of a single box ($z = 0.05$), and the right panel shows those over 10,000 sight lines stacked to $z = 1$. IllustrisTNG is shown in solid blue, SIMBA is plotted with the dashed orange line, and Astrid is shown by the dotted green line. The 1D distributions in the left panel correspond to the 2D maps in Figure 2. These confirm what we observe by eye in the maps: SIMBA displays a more uniform spread in baryons than IllustrisTNG and Astrid. The stacked sight-line distributions on the right amplify these differences.

## 4. Results

### 4.1. Cosmic DM

As our baseline, we start with the fiducial models for IllustrisTNG, SIMBA, and Astrid and compute the distribution of DM for 10,000 sight lines up to $z = 1$. The three distributions are shown in Figure 3, with SIMBA represented by orange, IllustrisTNG by blue, and Astrid by green. The mean DM value for the SIMBA fiducial model is $\langle DM \rangle = 1002.5 \pm 116.9$ cm$^{-3}$ pc, while the mean DM value for the IllustrisTNG fiducial run is $\langle DM \rangle = 1022.6 \pm 169.5$ cm$^{-3}$ pc, and for Astrid it is $\langle DM \rangle = 1019.9 \pm 217.7$ cm$^{-3}$ pc.

The mean and variance of the DM for each of the three models are compared to well-established ranges from the literature as an initial check on our methods and results. Before the discovery of FRBs, theoretical investigations assuming that all baryons are fully ionized hydrogen, homogeneously distributed, and in the IGM predicted that $\langle DM_{IGM}(z=1) \rangle \sim 1200$ cm$^{-3}$ pc (Ioka 2003; Inoue 2004). Since the discovery of FRBs in 2007 (Lorimer et al. 2007), numerous attempts have been made to quantify the distribution of DM. Medlock & Cen (2021), for instance, computed the distribution of 10,000 sight lines using the same light-cone building procedure with an $l = 50\ h^{-1}$ Mpc sized box using simulations described in Cen & Ostriker (2006) and Cen & Chisari (2011). For these simulations, at $z = 1$, the mean cosmic DM is given by $\langle DM \rangle = 919 \pm 202.3$ cm$^{-3}$ pc. Jaroszynski (2019) finds $\langle DM_{IGM}(z=1) \rangle = 905 \pm 115$ cm$^{-3}$ pc using Illustris, with simulation cubes of $l = 75\ h^{-1}$ Mpc. With IllustrisTNG-300, Zhang et al. (2021) obtain $\langle DM_{IGM}(z=1) \rangle \sim 892^{+721}_{-270}$ cm$^{-3}$ pc. Our results using CAMELS are consistent, within the errors, with the range of these previous simulation results. The predictions of Ioka (2003) and Inoue (2004) significantly overestimate the contribution of the IGM to the total DM due to simplifying assumptions. The Macquart relation takes a more realistic approach to the distribution of baryons and finds $\langle DM_{IGM}(z=1) \rangle \sim 1000$ cm$^{-3}$ pc, consistent with previous simulations and our results.

Now, we demonstrate that the differences in the DM distribution are due to the differences in subgrid implementations. Figure 2 reveals that DM maps alone show a notable difference in the baryon distribution of SIMBA, IllustrisTNG, and Astrid. This difference is even more evident in the distributions for the sight lines at $z = 1$. All three suites have similar mean DM values, with Astrid and IllustrisTNG being particularly close. However, the right panel of Figure 3 shows much more variance in the IllustrisTNG distribution than in SIMBA and even more in the Astrid distribution. The standard deviation in Astrid's distribution is nearly twice that of SIMBA's, with IllustrisTNG fitting neatly in between. SIMBA generally has higher gas temperatures in the IGM (Christiansen et al. 2020; Tillman et al. 2023a, 2023b) and stronger baryonic effects on the matter power spectrum than IllustrisTNG (Villaescusa-Navarro et al. 2021a; Delgado et al. 2023; Gebhardt et al. 2024; Pandey et al. 2023). This is because, in SIMBA, the AGN feedback turns on earlier and, in addition, incorporates the AGN jet mode feedback, which can push baryons to large scales (Davé et al. 2019; Borrow et al. 2020; Christiansen et al. 2020). Recently, the baryon spread metric for CAMELS was computed, finding that approximately 40% of baryons spread to distances greater than 1 Mpc in the SIMBA fiducial model, compared to 10% in IllustrisTNG and Astrid (Gebhardt et al. 2024). With stronger feedback and greater baryon spread, the baryon distribution over a box will be more uniform. This results in the narrower distributions for SIMBA in Figure 3.

Both IllustrisTNG and Astrid have a more condensed structure with large areas with very few baryons in between. Among the three suites, Astrid has the least impact on the matter power spectrum, as it has the mildest AGN feedback model among the three suites (Ni et al. 2023). On wide scales, IllustrisTNG and Astrid show relatively similar baryon spread distributions: 11% greater than 1 Mpc for IllustrisTNG and 7% greater than 1 Mpc for Astrid. However, compared to IllustrisTNG, Astrid displaces fewer baryons at intermediate distances (Gebhardt et al. 2024). This uneven distribution of baryons results in a wider distribution of DM values, as seen in Figure 3. Many sight lines will pass through these emptier regions with lower observed DM, while a small percentage will pass through highly dense filaments and halos with very high observed DM.

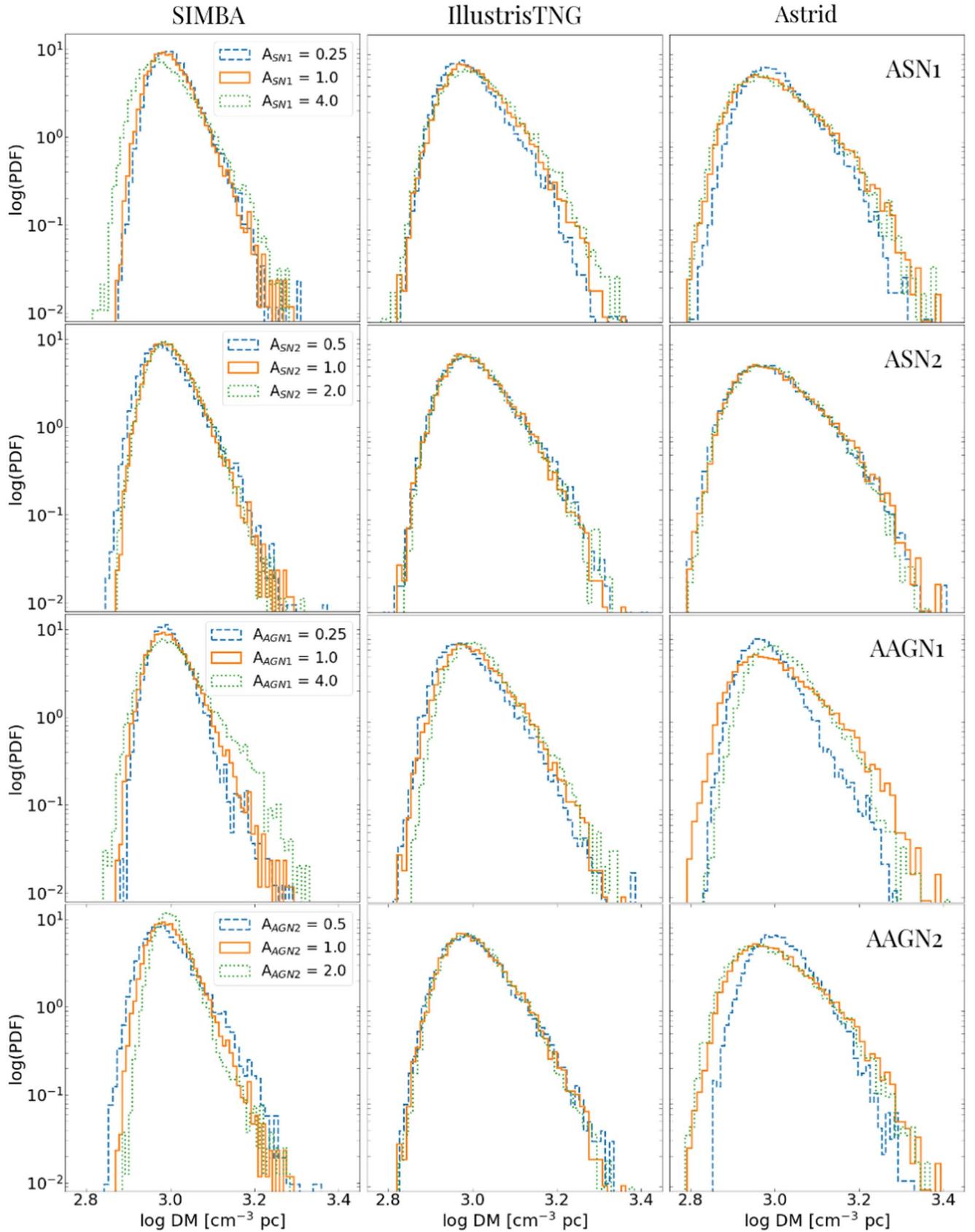

**Figure 4.** The complete set of DM distributions for sight lines to z = 1 from the 1P set for the three suites: SIMBA (left), IllustrisTNG (center), and Astrid (right). The astrophysical parameters are shown as follows: $A_{SN1}$ in the first row, $A_{SN2}$ in the second, $A_{AGN1}$ in the third, and $A_{AGN2}$ in the fourth. The fiducial distributions are plotted with a solid orange line. The minimum parameter value distributions are presented with dashed blue lines, and the maximum parameter value distributions are presented with dotted green lines. If we expect higher parameter values to equal higher feedback, then we would expect narrower distributions. This is not the case in all of these panels, indicating that something more complex is happening.

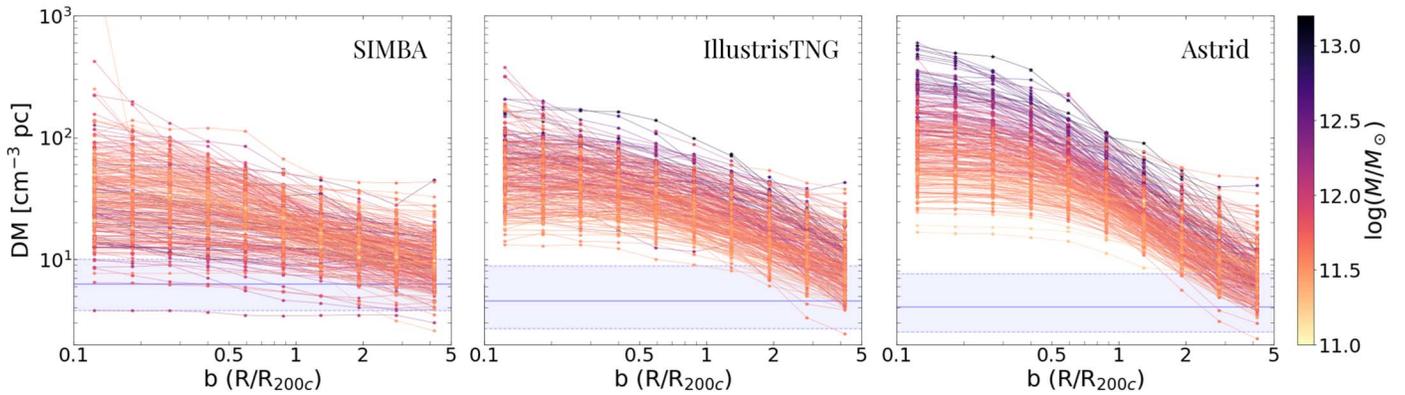

**Figure 5.** The halo DM profiles for the top 300 most massive halos in the fiducial run, $z = 0$ snapshot for SIMBA (left), IllustrisTNG (center), and Astrid (right). The impact parameter ($b = R/R_{200c}$) is plotted on the x-axis, and DM is plotted on the y-axis. The halo profiles are colored by the halo mass ($M_{200}$). The solid blue line indicates the mean DM of the IGM (defined as sight lines with an impact parameter greater than 3 for all halos), while the shaded blue region indicates the corresponding $1\sigma$ range of this background DM. This region marks where we transition from excess DM due to the halo to IGM. From left to right, the profiles display a general boost in DM, as expected, with a decreasing overall feedback strength. In addition, we observe an increasingly strong trend with halo mass from left to right.

We examine various simulations of the 1P set to understand how each of the six CAMELS parameters within a single subgrid model affects the observed DM distributions. For 10,000 sight lines up to $z \in [0.25, 0.5, 0.75, 1.0]$, we calculated the DM distributions for each simulation in the 1P set for the three suites. Figure 4 displays the complete set of distributions for the sight lines at $z = 1$. The suites are presented in the columns from left to right: SIMBA, IllustrisTNG, and Astrid. Astrophysical parameters are presented from top to bottom: $A_{SN1}$, $A_{SN2}$, $A_{AGN1}$, and $A_{AGN2}$. The fiducial model is plotted in orange, the lowest value is in blue, and the highest value is in green.

Generally, the greatest difference arises between suites, following the trends observed in Figure 3. Within the suites, we see that certain subgrid models exhibit more drastically modified distributions when an astrophysical parameter is varied. Generally, IllustrisTNG appears to be most robust to changes in parameter values. However, the Astrid AGN parameters show the most significant difference in the distributions. The two bottom right panels show the distributions of varying $A_{AGN1}$ and $A_{AGN2}$ for Astrid. A lower value of $A_{AGN1}$ results in a narrower distribution, indicating a more uniform baryon distribution. The maximum value of $A_{AGN1}$ also has a similar effect but not to the same extent. When $A_{AGN2}$ for Astrid is varied, the fiducial and maximum $A_{AGN2}$ distributions are similar, while the minimum $A_{AGN2}$ distribution is significantly narrower. If increasing both AGN parameters led to stronger feedback overall, we would expect a different trend, where the higher-value parameter distributions should be narrower. This suggests that there is significant complexity in how the AGN parameters interact with other properties of halos and galaxies, which we will discuss in more detail in Section 4.3.

In the top left panel, we analyze the distributions of SIMBA while varying $A_{SN1}$. The left column, third row panel shows the distributions resulting from the variation of $A_{AGN1}$ in SIMBA. The minimum parameter value distribution and the fiducial in both panels are similar, but the maximum parameter value distribution is significantly wider. For $A_{SN1} = 4.0$, the distribution is enhanced at the lower end of the DM sight line, whereas for $A_{AGN1} = 4.0$, the distribution is enhanced at the upper end of the DM sight line. These trends are the opposite of what we expect if increasing a given feedback parameter corresponds to greater overall feedback.

### 4.2. Contribution of the CGM to DM: A Halocentric Approach

We now turn our attention from the impact of feedback on the overall distribution of DM to its impact on individual halos, specifically the CGM. We plot the impact parameter versus the median DM in Figures 5 and 6. Lastly, we calculate $\langle DM_{IGM} \rangle$ by computing the median DM for cells with $b > 3$, which we will consider to be IGM. Figure 5 shows the DM profiles of the 300 most massive halos in the fiducial simulation for SIMBA, IllustrisTNG, and Astrid at snapshot $z = 0$, plotted against the impact parameter ($b = R/R_{200c}$). From right to left, we see an increase in suppression of halo DM profiles, with SIMBA exhibiting the strongest suppression due to its stronger AGN feedback, which evacuates more of the baryons from halos. This is analogous to the impact of feedback on baryon spread discussed in Section 4.1. Therefore, we expect that SIMBA halo profiles are more suppressed than IllustrisTNG and Astrid, as a significant portion of the baryons are pushed out beyond $R_{200c}$.

In Figure 5, we also observe that excess DM relative to the IGM persists well beyond $R_{200c}$ in IllustrisTNG and Astrid and, to a lesser extent, in SIMBA. This might seem initially surprising, as we know that SIMBA pushes baryons out further. However, this shows the contrast between halos and the background DM. As the baryon distribution in SIMBA is more uniform with higher amplitude, it makes sense that the profiles would reach background DM at a smaller radius. In addition, the SIMBA profiles are flatter and do not have the steeper profile with the impact parameter, as Astrid does. In fact, the jet mode of AGN feedback plays a large role in moving baryons from the CGM to the IGM. For instance, deactivating jet mode AGN feedback in SIMBA results in a ~20% drop in the baryon fraction of the IGM (Khrykin et al. 2024).

In addition, there is significant scatter in the profiles. Astrid has a trend between halo mass and excess DM, where lower-mass halos exhibit lower DM profiles. This trend is also present in IllustrisTNG, although to a lesser degree. SIMBA does not show any noticeable mass trend. It is expected that larger halos would hold more baryon content, which makes the trend seen in Astrid and IllustrisTNG unsurprising.

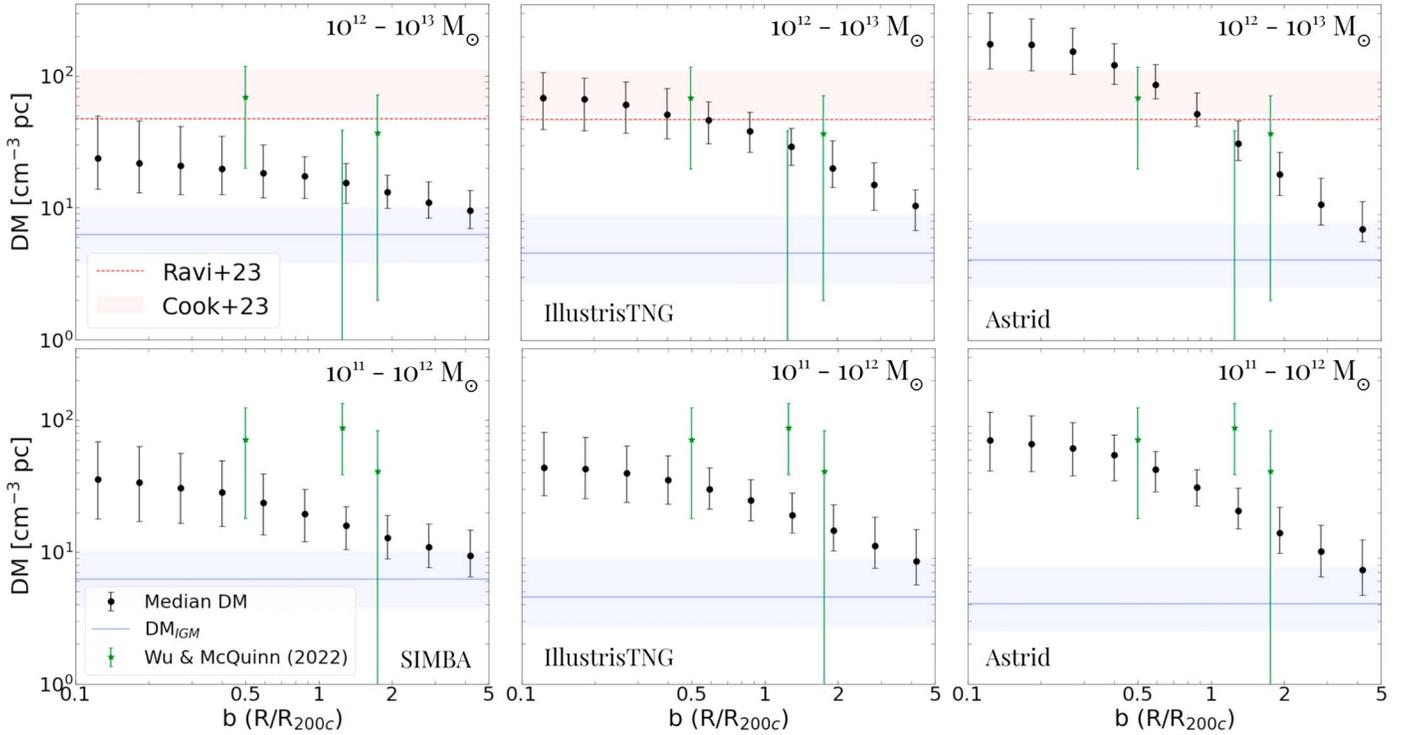

**Figure 6.** The median halo profiles for the top 300 most massive halos in the fiducial run, $z = 0$ snapshot for SIMBA (left column), IllustrisTNG (middle column), and Astrid (right column). The impact parameter ($b = R/R_{200c}$) is plotted on the x-axis and the DM on the y-axis. The profiles are binned by mass, with the bottom row showing halos of mass $10^{11}$–$10^{12}\ M_\odot$ and the top row showing halos of mass $10^{12}$–$10^{13}\ M_\odot$. As in Figure 5, the solid blue line indicates the mean DM of the IGM (defined as sight lines with an impact parameter greater than 3 for all halos), while the shaded blue region indicates the $1\sigma$ range. Green stars indicate excess DM as measured by Wu & McQuinn (2023) within the appropriate halo mass bins. The red lines show the upper limit of the observed Milky Way CGM DM observed by Ravi et al. (2023; dashed line) and Cook et al. (2023; shaded region).

To examine the mass trends more closely, we plot the median halo profiles sorted by mass in Figure 6. The first bin includes halos with a mass of $M_{halo} = 10^{11}$–$10^{12}\ M_\odot$, and the second bin includes halos with $M_{halo} = 10^{12}$–$10^{13}\ M_\odot$. The first bin has the most halos, as the halo mass function increases with less massive halos (see the mass distribution in Figure 1). These profiles illustrate the relationship between the halo mass and excess DM. The data show that in Astrid and IllustrisTNG, larger-mass halos have a greater amount of excess DM. However, the trend is the opposite in SIMBA, with higher-mass halos having slightly suppressed profiles compared to lower-mass halos. The SIMBA profiles are significantly more suppressed than those of Astrid and, to a lesser extent, IllustrisTNG. For the lower mass bin, the peaks of the IllustrisTNG halo profiles are approximately 1.25 times greater than those of SIMBA, whereas Astrid's peaks are roughly twice as high. The difference is even more pronounced in the higher mass bin, with IllustrisTNG's peaks being approximately 3 times higher than SIMBA's and Astrid's peaks being around 7 times higher.

Over the past few years, there have been exciting developments in measuring CGM with FRBs as the number of localized FRBs has increased. In Figure 6, we compare our predictions against three recent measurements and the halo profiles. Recently, upper limits on the contribution of the Milky Way halo to the DM have been placed via observations (Cook et al. 2023; Ravi et al. 2023). To obtain these constraints, the FRB must be well localized to identify the host halo and distance. $DM_{MW,Halo}$ is estimated by subtracting the other contributions from the measured DM. The Macquart relation or another theoretical framework is used to estimate $DM_{IGM}$, and a nominal value is estimated for $DM_{Host}$. Nearby pulsars are used to estimate $DM_{MW,ISM}$ from either the NE2001 or YMW16 model of the Galactic ionized ISM distribution. Thus, the observational constraints are limited due to the uncertainty in each of these estimated components. Using FRB 20220319D, discovered and localized by DSA-110 at a distance of $\sim 50$ Mpc, Ravi et al. (2023) report a limit of $DM_{MW,Halo} \leqslant 28.7$ or $DM_{MW,Halo} \leqslant 47.3$ cm$^{-3}$ pc, depending on which nearby pulsar is used to estimate $DM_{ISM}$ and assuming that $DM_{IGM} = 7$ and $DM_{Host} = 10$ cm$^{-3}$ pc. This result implies that the CGM mass is less than $10^{11}\ M_\odot$ and that the Milky Way contains less than 60% of the cosmological baryon density, supporting the picture that feedback drives baryons out. Likewise, using the CHIME FRB population, Cook et al. (2023) place a limit of $DM_{MW,Halo} = 52$–$111$ cm$^{-3}$ pc, depending on Galactic latitude. This range is found by fitting various models to the Galactic latitude and $DM_{MW,Halo}$ of 93 CHIME FRB sources.

As these upper limits apply only to the Milky Way, we compared them with the halos in the $10^{12}$–$10^{13}\ M_\odot$ mass bin. SIMBA's halo profiles remain below this upper limit, while portions of the Astrid and IllustrisTNG profiles exceed the estimated upper limits. However, since a significant portion of the halos included in these median profiles have masses above that of the Milky Way, we cannot expect these upper limits to apply. We cannot make a direct comparison with these observations, but we can see that our halo profiles correspond to the expected Milky Way profile within an order of magnitude.

Recently, Connor & Ravi (2022) analyzed excess DM from halos beyond the Milky Way using the CHIME FRB sample

(CHIME/FRB Collaboration et al 2021). To perform this analysis, excess DM is estimated by identifying FRBs in a larger sample that intersect halos and comparing the mean DM of these to the mean DM of the entire FRB population. Building on this analysis, Wu & McQuinn (2023) calculated the excess DM profiles of halos with two mass bins of $M_{\rm halo} = 10^{11}$–$10^{12}$ and $M_{\rm halo} = 10^{12}$–$10^{13}$ $M_\odot$ and impact parameter bins of [0, 1], [1, 1.5], and [1.5, 2] $R/R_{200c}$. This analysis also includes a weighting scheme for calculating excess DM to mitigate the effects of the small sample size of FRBs that intersect halos. Even so, the low number of well-localized FRBs leads to large uncertainty in these constraints. Figure 6 plots these measurements in green. Our halo profiles fit roughly within the measured $1\sigma$ range.

### 4.3. Baryonic Feedback and the F Parameter

In this work, we use $F$ to measure the fluctuation in DM caused by feedback and large-scale structure. We define $F$ in Equation (8) and discuss its origin in Section 3.5. To assess the impact of varying parameters, we computed $F$ for each simulation in the 1P sets of SIMBA, IllustrisTNG, and Astrid. Figure 7 shows the results, with cosmological parameters in the left column, supernova feedback parameters in the middle column, and AGN feedback parameters in the right column. Recall that a higher value for $F$ corresponds to a weaker feedback, and a lower value closer to 0 corresponds to stronger feedback. Our analysis shows that SIMBA produces the lowest $F$ values, indicating stronger feedback, while Astrid has the highest $F$ values, indicating weaker feedback. IllustrisTNG falls into the intermediate range and displays the least variance between the 1P set simulations. This is consistent with our previous findings, where SIMBA has the most uniform distribution of baryons, Astrid has the most condensed structure, and IllustrisTNG falls in between.

Intuitively, we would expect that increasing a feedback parameter would increase the overall feedback. However, this is not always the case. In fact, according to Figure 7 and earlier sections, the $F$ value exhibits inverse behavior within each pair of astrophysical parameters. For instance, in IllustrisTNG, increasing the value of $A_{\rm SN1}$ results in an increase in $F$, while increasing $A_{\rm SN2}$ leads to a decrease in $F$, as expected. In the upcoming sections, we will examine the behavior of each subgrid model's six parameters.

#### 4.3.1. Cosmological Parameters

There is a clear positive correlation between $\Omega_M$ and $\sigma_8$ with $F$ in the three suites. As $\Omega_M$ increases, the effect of gravity becomes more dominant, resulting in more collapsed structures and a higher $F$. As $\sigma_8$ increases, the scale we need to smooth over to get a homogeneous Universe increases. This means that we should observe a more differentiated structure, leading to a higher $F$. The two cosmological parameters have the greatest impact on the large-scale end of the matter power spectrum (Ni et al. 2023). In addition, increasing both cosmological parameters promotes the earlier formation of more massive halos, galaxies, and black holes, causing nonlinear effects on the matter power spectrum earlier.

#### 4.3.2. Supernova Feedback Parameters

We have noticed a consistent pattern with both $A_{\rm SN1}$ and $A_{\rm SN2}$ across all three suites. When we increase the value of $A_{\rm SN2}$, the parameter $F$ decreases. $A_{\rm SN2}$ has a direct influence on the speed of galactic winds that are produced by stellar feedback in all three suites. As the speed of these winds increases, they become more effective in removing gas from halos, resulting in a more uniform distribution of baryons throughout the box, ultimately leading to a lower value of $F$. This is in line with the discoveries made by Nicola et al. (2022), who found that as $A_{\rm SN2}$ increases in the IllustrisTNG 1P set, the proportion of cosmic gas in halos ($f_{\rm cosmic}$) decreases. Similar trends of the increasing halo baryon fraction as a function of $A_{\rm SN2}$ were presented in Delgado et al. (2023).

When the value of $A_{\rm SN1}$ increases, it also causes $F$ to increase. This indicates that the distribution of baryons becomes less uniform and is consistent with the finding that increasing $A_{\rm SN1}$ leads to a higher halo baryon fraction (Delgado et al. 2023). It has been observed that increasing $A_{\rm SN1}$, which refers to galactic winds, has an inhibitory effect on massive galaxy formation and thus black hole growth (Tillman et al. 2023b). Studies have shown that as $A_{\rm SN1}$ increases, the abundance of galaxies at all masses is suppressed in all three suites, and the black hole mass is also greatly suppressed in SIMBA. This leads to a decrease in overall feedback since AGN feedback is expected to dominate the total feedback energy. Our ongoing research (I. Medlock et al. 2024, in preparation) has found that an increase in $A_{\rm SN1}$ in the 1P set in IllustrisTNG is associated with lower black hole masses and lower total feedback energy. However, increasing $A_{\rm SN2}$ has no noticeable impact on black hole size and does not inhibit AGN feedback, which is consistent with our results. Another study by Gebhardt et al. (2024) has also shown that increasing $A_{\rm SN2}$ decreases the spread of baryons in all three suites.

#### 4.3.3. AGN Feedback

The impact of changing the AGN parameters differs depending on the subgrid model used. Astrid, in particular, exhibits unique behavior compared to the other two suites. In Astrid, increasing $A_{\rm AGN1}$ leads to an initial increase in $F$, which then peaks near the fiducial value before decreasing. In contrast, for both SIMBA and IllustrisTNG, $F$ increases as the parameter increases. However, $A_{\rm AGN1}$ has a more limited effect due to the small volume of the CAMELS boxes, which inhibits the formation of the most massive structures that would display stronger AGN jet feedback. SIMBA has a lower black hole mass threshold for turning on the AGN jet mode feedback, which results in a stronger effect than TNG. The jet mode of AGN feedback in SIMBA has the greatest effect on cold gas (Davé et al. 2020). Increasing $A_{\rm AGN1}$ in SIMBA, which controls momentum flux in jet mode black hole feedback, leads to a suppression of higher-mass black holes and galaxies (Ni et al. 2023), resulting in an overall decrease in feedback.

In Astrid, the value of $F$ increases as $A_{\rm AGN2}$ increases. This is because strong AGN thermal feedback in Astrid limits the growth of black holes and ultimately restricts the overall amount of feedback (Ni et al. 2023). However, in SIMBA and IllustrisTNG, $F$ decreases as $A_{\rm AGN2}$ increases. This is because $A_{\rm AGN2}$ controls the jet mode, which is positively correlated with feedback, resulting in a decrease in $F$ in both SIMBA and IllustrisTNG.

#### 4.3.4. Comparison to Observations

Recent work has measured $F$ observationally. Macquart et al. (2020) reported $F = 0.23^{+0.27}_{-0.12}$ with seven FRBs and adopted a prior on $\Omega_b h$ based on studies of the cosmic microwave background, Big Bang nucleosynthesis, and supernovae. Better constraints can be obtained with the growing detection of FRBs and localizations. Recently, Baptista et al. (2024) measured $F = 0.33^{+0.27}_{-0.11}$ with 78 FRBs (21 with redshifts) and assumed a uniform prior on the Hubble parameter ($h \in [0.67, 0.73]$). This result aligns with the $F = 0.44^{+0.31}_{-0.13}$ calculated for IllustrisTNG, extrapolated from the results of Zhang et al. (2021), with a prior of $h = 0.67$, as in CAMELS. Crucially, a lower limit of $F > 0.128$ with 99.7% confidence was measured, which is shown in Figure 7 by the dashed black line.

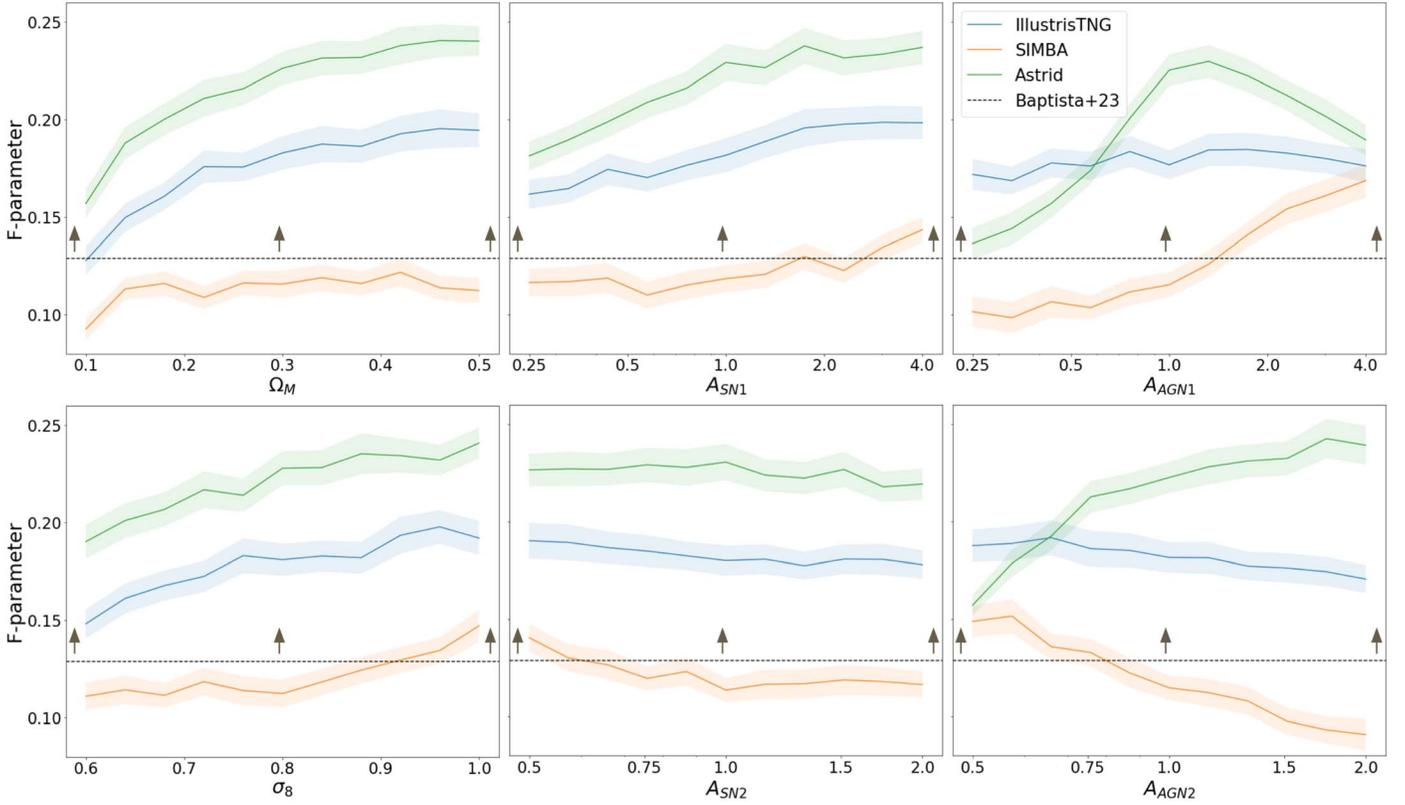

**Figure 7.** The $F$ parameter for each simulation in the 1P set for the three suites. Each panel shows the variation of one of the six parameters. On the $x$-axis, we plot the value of the varying parameter, with the fiducial value at the center, against $F$ on the $y$-axis. The $1\sigma$ error is indicated by the shaded region surrounding each line. The results for IllustrisTNG are shown in blue, SIMBA in orange, and Astrid in green. We compare with the observed $F$ measured by Baptista et al. (2024), who found a lower limit of $F = 0.128$ with 99.7% confidence. Most SIMBA runs, including the fiducial model, are in tension with the measured lower limit.

We found $F = 0.182^{+0.018}_{-0.017}$, $F = 0.119 \pm 0.014$, and $F = 0.229 \pm 0.019$ for the fiducial models of IllustrisTNG, SIMBA, and Astrid, respectively. The values for IllustrisTNG and Astrid fall within the $1\sigma$ range of these measurements, while SIMBA falls well below. All of the Astrid and IllustrisTNG models fall above this lower limit. Most SIMBA models, including the fiducial model, fall below this lower limit, creating tension. However, we expect that the $F$ values measured here are biased to the lower range due to the issues with cosmic variance in CAMELS. This will be discussed in more detail in Section 5.

## 5. Discussion

### 5.1. Limitations

Our simulations are limited in two ways. First, the size of the CAMELS box (25 $h^{-1}$ Mpc) is too small to account for the large-scale cosmic variance. This means that it cannot capture long-wavelength modes that are crucial for the formation of more massive objects, such as galaxy clusters. Consequently, the matter power spectrum is not properly normalized on all scales (Villaescusa-Navarro et al. 2021a). The distribution of dark matter is highly sensitive to cosmic variance. For example, examining the contribution of large-scale structures to dark matter, it was found that a significant portion of the scatter in dark matter sight-line distributions is due to both halos and filamentary matter (Walker et al. 2024). The CAMELS boxes do not contain any halos with $M_{\rm halo} > 10^{13.3}\ M_\odot$, so they cannot capture the impact of these large halos that are expected to contribute significantly to DM. Furthermore, from calculations of dark matter distributions using simulations with varying box sizes ($L = 25, 50,$ and $100\ h^{-1}$ Mpc), it was discovered that simulations with box sizes around $100\ h^{-1}$ Mpc are necessary to capture the effect of the lognormal matter density profile on the cosmic dark matter distribution (Batten et al. 2021). Consequently, the CAMELS project simulations cannot fully capture the distribution of dark matter and the impact of large-scale structures. Thus, our calculated values for $F$ are biased toward the low end, and we cannot presently rule out any models. We roughly estimate that a correction of an increase of ~50% is needed for the CAMELS $F$ values based on the convergence study of Batten et al. (2021). To estimate this correction, we fit a cubic spline to the values of Figure C2 of their paper, which plots $\sigma_{\rm DM}$ as a function of $z$ for Eagle boxes of 25, 50, and 100 cMpc, to measure the expected increase in $\sigma_{\rm DM}$ from a box size of 25 $h^{-1}$ Mpc to 100 cMpc. The degree of bias on $F$ needs to be further quantified to determine if any models in the CAMELS project are in tension with the lower limits of $F$ calculated with observations.

Second, due to the high computational costs involved, CAMELS can only vary six parameters: two cosmological parameters and four astrophysical ones. Unfortunately, this means that these parameters alone cannot fully capture the complex dynamics involved in cosmology and galaxy formation and evolution. Additionally, the four feedback parameters used in CAMELS may not be the most representative. For instance, our findings show that increasing certain parameters does not necessarily lead to higher overall feedback. There are likely complex effects and interactions between feedback parameters that are not captured in the current CAMELS.

### 5.2. Future Directions

To address concerns about the accuracy of our results, we could repeat our analysis in a suite with a larger box size. Unfortunately, these simulations are currently unavailable. However, the CAMELS project is in the process of running a new set of simulations with box sizes of 50 $h^{-1}$ Mpc, which will allow us to test the robustness of our findings against more precise simulations that capture cosmic variance. Meanwhile, we have explored a few options to address this concern. One approach is to compute a "correction factor" to account for the lack of variance in the CAMELS boxes. This involves computing the fiducial model's $F$ out to several box sizes until it converges and generates a numerical factor. Previous work has used various strategies to mitigate this issue, such as employing parameters that encode the distribution of halo masses to reduce the effect on neural networks trained to constrain parameters using the electron density power spectrum (Nicola et al. 2022) or scaling the CAMELS halo mass functions to those of large-volume simulations and applying them to feedback-constraining spectral distortion measures (Thiele et al. 2022). In essence, while CAMELS is not designed to provide absolute estimates due to the small boxes, it does provide relative differences between runs. Thus, we can use CAMELS to calibrate relative differences and then anchor the results to larger boxes (such as TNG300) to link to the absolute scale.

To address concerns involving the limited parameter space of the CAMELS project, there are a couple of steps forward. First, recently, the CAMELS collaboration expanded the parameter space with the SB-28 and 1P extended sets, which include 28 parameters, presented in Ni et al. (2022). We plan to repeat our analysis using some of these new parameters. Second, we are working on directly quantifying the feedback energetics in CAMELS to provide a more detailed explanation for our results. Additionally, we can repeat the analysis on the Latin hypercube and SB-28 CAMELS sets. Simulations in these sets take into account the effects of multiple parameters and may highlight some of the complicated interrelations between them that cannot be studied with the 1P set. Finally, we plan to extend our comparisons of subgrid models to include additional suites such as Ramses, Enzo, and Magneticum as they become available.

In terms of observations, it is important to prepare our theoretical predictions for the upcoming increase in FRB localizations. It is expected that with 100 well-localized FRBs, we can place a significant limit on $F$ (Baptista et al. 2024). In addition, it is predicted that we only need an order of 10–100 localized FRBs to constrain CGM density profiles (Ravi 2019). It is expected that with $10^3$–$10^4$ FRB detections, we can statistically detect gas densities at various impact parameters of the CGM of the intervening galaxies (Ravi et al. 2019). In our analysis, we are modeling the DM sight lines without the host and MW contributions to the DM ($DM_{Host} + DM_{MW}$). To compare directly with observations, one needs to include the most accurate estimates for these. In addition, one must incorporate any systematic effects that may arise in observing. The constraining power of FRBs can be improved by combining them with other multiwavelength probes as well (Battaglia et al. 2019).

### 6. Conclusions

Complex baryonic feedback processes leave imprints in the baryon distribution, including in the CGM and IGM. These processes are crucial components in galaxy formation and evolution. We examine CAMELS to explore the impact of baryonic feedback on the DM of FRBs, focusing on the CGM. We analyze three simulation suites in the CAMELS project: IllustrisTNG, SIMBA, and Astrid. For each, we examine the effect of varying six parameters (two cosmological and four pertaining to feedback). For each of these parameters, we analyze 11 simulations with varied values for a total of 66 simulations per suite and 183 simulations all together. We calculate both IGM and CGM statistics, allowing direct comparison to recent observational data.

Here we summarize the main findings of our study.

1. In the fiducial runs, SIMBA displays the greatest impact from feedback, with a more uniform baryon distribution than IllustrisTNG and Astrid. This is evident in the DM maps within a single box but more noticeable in the DM distributions across cosmological redshifts. The SIMBA runs have a narrower DM distribution with less variation, whereas IllustrisTNG has a wider distribution that includes both high-DM sight lines passing through dense filaments and low-DM sight lines passing through voids. Astrid displays even more variation, indicating that it is the least affected by feedback.
2. SIMBA is more affected by AGN feedback than IllustrisTNG and Astrid, possibly because its black hole mass threshold for feedback activation is lower. $A_{SN1}$ has a strong effect on Astrid, where greater supernova feedback leads to a decrease in the growth of black holes and galaxies, resulting in overall feedback suppression.
3. In halos, Astrid and IllustrisTNG predict a higher amount of DM (by a factor of ∼2–7) compared to SIMBA, which extends to several virial radii. This is because strong feedback from the AGN jet causes halos to be evacuated in SIMBA runs. Additionally, Astrid and IllustrisTNG predict an excess of DM beyond a factor of 5 (i.e., 5 times the virial size of a halo) compared to SIMBA.
4. We find a significant correlation between mass and the normalized DM profiles in Astrid, with a smaller correlation in IllustrisTNG. As the mass of the halos increases, we see an increase in DM. However, in SIMBA, there is a weak inverse trend, where lower-mass halos exhibit a slight increase in DM.
5. We calculate the $F$ parameter for all runs in the 1P set of SIMBA, IllustrisTNG, and Astrid. Upon comparing our findings with observations from Baptista et al. (2024), we find that a substantial portion of the CAMELS-SIMBA 1P set, including the fiducial model, falls beyond the confidence lower limit of 99.7%. However, we cannot account for cosmic variance, which will increase the calculated *F* parameter (see Section 5). Therefore, it is too early to tell if there is a tension. Thus, it is necessary to repeat this analysis either with a correction factor for the *F* parameter or, ideally, on larger boxes that can properly capture cosmic variance.

The discovery of FRBs is an exciting development for studying the distribution of baryons in galaxies and intergalactic space. With the help of advanced technology and increasing computational power, we can now make more accurate predictions and detect more FRBs than ever before. For example, facilities such as DSA-2000 are expected to detect and locate up to 10,000 FRBs each year, providing ample opportunity to study the IGM (Ravi et al. 2019). Efforts such as the FLIMFLAM survey will constrain the distribution of cosmic baryons and elucidate the effects of foreground halos and their CGMs (Lee et al. 2023; Simha et al. 2023). Furthermore, combining FRBs with other probes, such as the SZ effect, can help us better understand the distribution of baryons in the Universe (e.g., Battaglia et al. 2019; Pandey et al. 2023). These tools will allow us to make robust measurements of the *F* parameter and the feedback effects on the CGM in the coming years.

## Acknowledgments

The authors acknowledge constructive comments and suggestions from the anonymous referee and helpful discussions with Jason Prochaska, Marcin Glowacki, Calvin Leung, and Jay Baptista. The CAMELS were performed on the supercomputing facilities of the Flatiron Institute, which is supported by the Simons Foundation. This work is supported by NSF grant AST 2206055 and the Yale Center for Research Computing facilities and staff. I.M. would like to acknowledge support from the Yale Graduate School of Arts & Sciences through the Dean's Emerging Scholars Research Award. P.S. acknowledges support from the YCAA Prize Postdoctoral Fellowship. D.A.A. acknowledges the support of NSF grants AST-2009687 and AST-2108944, CXO grant TM2-23006X, JWST grant GO-01712.009-A, Simons Foundation Award CCA-1018464, and Cottrell Scholar Award CS-CSA-2023-028 by the Research Corporation for Science Advancement.

## ORCID iDs

Daisuke Nagai  https://orcid.org/0000-0002-6766-5942
Benjamin Oppenheimer  https://orcid.org/0000-0002-3391-2116
Daniel Anglés-Alcázar  https://orcid.org/0000-0001-5769-4945